\documentclass[preprint,aps,prc,superscriptaddress,showpacs,nofootinbib,floatfix]{revtex4}
\usepackage{amssymb}
\usepackage{graphicx,color}
\usepackage{bm}

\usepackage{graphicx} \usepackage{dcolumn} \usepackage{bm}

\newcommand{\beq}{\begin{equation}}
\newcommand{\eeq}[1]{\label{#1}\end{equation}}
\newcommand{\beqar}{\begin{eqnarray}}
\newcommand{\eeqar}[1]{\label{#1}\end{eqnarray}}

\newcommand{\bmath}{\begin{displaymath}}\newcommand{\emath}{\end{displaymath}}\newcommand{\bitem}{\begin{itemize}}\newcommand{\eitem}{\end{itemize}}
 
\begin{document}

\title{\Large \bf 
Predictions for $p+{\rm Pb}$ at 4.4$A$ TeV to test 
initial state nuclear shadowing at the Large Hadron Collider.}

\newcommand{\columbia}{Columbia University, New York, N.Y. 10027, USA}

\newcommand{\mcgill}{McGill University, Montreal, H3A 2T8, Canada}

\newcommand{\kfki}{WIGNER RCP, Institute for Particle and Nuclear Physics
 P.O.Box 49, Budapest, 1525, Hungary}

\author{~G.~G.~Barnaf\"oldi} \affiliation{\kfki}
\author{~J. Barrette} \affiliation{\mcgill}
\author{~M.~Gyulassy} \affiliation{\columbia}\affiliation{\kfki}
\author{~P.~Levai} \affiliation{\kfki}
\author{~V.~Topor~Pop} \affiliation{\mcgill}

\date{January 23, 2012}

\begin{abstract}

Collinear factorized perturbative QCD model predictions
are compared for $p+{\rm Pb}$ at 4.4$A$ TeV to test 
nuclear shadowing of parton distribution at the Large Hadron Collider (LHC).  
The nuclear modification factor (NMF), 
$R_{p{\rm Pb}}(y=0,p_T<20\;{\rm GeV}/{\it c}) = dn_{p{\rm Pb}}
/(N_{\rm coll}(b)dn_{pp})$, 
is computed with electron-nucleus  
($e+A$) global fit with different nuclear shadow distributions
and compared to fixed $Q^2$ shadow ansatz used in 
Monte Carlo  Heavy Ion Jet Interacting Generator ({\small HIJING}) type models.
Due to rapid DGLAP reduction of shadowing with increasing $Q^2$ used
in $e+A$ global fit, our results confirm   
that no significant initial state suppression
is expected ($R_{p{\rm Pb}} (p_T) = 1 \pm 0.1$) in the $p_T$ range 
5 to 20 GeV/{\it c}. In contrast, the fixed $Q^2$
shadowing models assumed in 
{\small HIJING} type models predict in the above $p_T$ range 
a sizable suppression, $R_{p{\rm Pb}} (p_T) = 0.6-0.7$ at mid-pseudorapidity 
that is similar to the color glass condensate (CGC) model predictions.
For central ($N_{\rm coll} = 12$) $p+{\rm Pb}$ collisions and at 
forward pseudorapidity ($\eta = 6$) the {\small HIJING} type models predict   
smaller values of nuclear modification factors ($R_{p{\rm Pb}}(p_T)$) 
than in minimum bias events at mid-pseudorapidity ($\eta = 0$). 
Observation of $R_{p{\rm Pb}}(p_T= 5-20 \,\,{\rm GeV}/{\it c}) 
\stackrel{<}{\sim} 0.6$ for minimum bias $p+A$ collisions  
would pose a serious difficulty for separating initial 
from final state interactions in Pb+Pb collisions at LHC energies. 


\end{abstract}


\maketitle


\section{Introduction}

In this paper we compare predictions for moderate $p_T<20$ GeV/{\it c} 
observables in $p+{\rm Pb}$ reactions at 4.4$A$ TeV 
at the Large Hadron Collider (LHC), that will help to
discriminate between models of initial conditions 
assumed in Pb+Pb collisions at 2.76$A$ TeV.  The possibility that
the first data on $p+{\rm Pb}$ 
may be taken soon, with a potential high physics payoff
motivates this paper. All models details are extensively discussed
in the literature 
and we focus only on the updated nuclear modification factor, 
$R_{p{\rm Pb}}(\eta,p_T,b)= dn_{p{\rm Pb}}/ 
(N_{\rm coll}(b) dn_{pp})$,
predictions testable with a short 4.4$A$ TeV run.
In minimum bias (MB) and central ($0-20\%$) $p+{\rm Pb}$ collisions 
the average number of binary 
nucleon-nucleon ($NN$) interactions (with an inelastic cross section 
$\sigma_{NN}^{\rm in} \approx 65$ mb) is $N_{\rm coll}^{\rm MB} \approx 7$ 
and $N_{\rm coll}^{\rm Cen} \approx 12$ respectively. 

This control experiment has long been anticipated to play
a decisive  role in  helping to deconvolute initial and final state
interaction effects in Pb+Pb reactions at the LHC. 
The  $d+{\rm Au}$ control experiment at 0.2$A$ TeV played a similar critical
role for Au+Au at the Relativistic Heavy Ion Collider
(RHIC) in 2004 \cite{RHIC,Gyulassy:2004zy}. The importance of $p+{\rm Pb}$ was 
also emphasized in the 2007 
Last Call for LHC compilation of predictions~\cite{armesto2_08} and many
other works ~\cite{Eskola:1998df,EPS08,EPS09,Accardi:2004gp,Eskola:2010jh,pALHC11,Wiedemann10}. 

The open problem after the first very successful 
LHC heavy ion run in 2010 \cite{QM11} remains
how to deconvolute nuclear modification effects  due to initial state 
and final state effects. Without a clear calibration of
the magnitude of initial state suppression of the incident
nuclear partonic flux it is not possible 
to draw firm conclusions
about the properties of the  quark gluon plasma (QGP) phase of  
matter produced at the LHC. 
At RHIC the same problem was resolved at mid-rapidity  by the observation 
of no appreciable nuclear modification 
in $d+{\rm Au}$ control experiments in 2003~\cite{RHIC,Gyulassy:2004zy} (see
also Figure~\ref{fig:fig3} below). 
An approximately factor of four  
suppression of moderate $p_T$ mid-rapidity pions observed in Au+Au at RHIC
could then be interpreted as due to final state jet energy loss 
 in a high opacity QGP produced in central Au+Au collisions at  RHIC.

At the LHC the initial flux is much more uncertain than at RHIC
because of the higher density of partons at an order of magnitude
smaller fractional momenta $x=2p_T/ \sqrt{s}< 10^{-3}$. 
At high initial densities all models predict a  breakdown 
of additivity of the nuclear parton distribution
functions (nPDF). The magnitude of the  breakdown however 
varies greatly in the literature 
in both collinear factorized approaches and $k_T$ factorized 
parton saturation model approaches
~\cite{armesto2_08,Eskola:1998df,EPS08,Accardi:2004gp,Eskola:2010jh,pALHC11,Wiedemann10}.
Therefore, even a rough first experimental constraint
from $p+Pb$ interactions would have high impact on the development
of nuclear collision modeling. 

\section{Nuclear shadowing and jet quenching at LHC energies}

Nuclear shadowing of quark and gluon nPDFs 
at large  $x > 0.01 $ and moderate $Q^2$ 
is well constrained from $e+A$ and lower energy $p+A$ data.
Global fit parametrizations of the nuclear parton distribution
functions (nPDF) are available~\cite{EPS08,Eskola:2010jh,pALHC11}. 
DGLAP evolution \cite{parisi_77} to higher $Q^2$ predicts a rapid 
reduction of shadowing
effects and therefore only modest modifications of 
$R_{p{\rm Pb}} (p_T) = 1\pm 0.1$ for $p_T>5$ GeV/c have been 
predicted~\cite{pALHC11,Wiedemann10,Levai11}.
As emphasized in~\cite{Wiedemann10}, any observed significant
 modification from unity would be inconsistent with most current nPDFs
and therefore pose a severe challenge to conventional  collinear factorized
QCD approximation to high-$p_T$ processes not only in $p+{\rm Pb}$ 
but even more so
in Pb+Pb collisions. We continue here to investigate this central thesis.

At RHIC there is clear
evidence at high rapidities, where small fractional parton momenta 
$x\sim 10^{-3}$ similar to central $p+{\rm Pb}$ are probed, that
binary collision scaling of 
collinear factorization breaks down. Color Glass Condensate (CGC)
$k_T$ factorization models have been developed to explain these 
deviations~\cite{Kharzeev:2003wz, Kharzeev:2004if} and nearly identical
$R_{pA}\approx 0.7 \pm 0.1$ nuclear modifications factors were predicted 
in Ref.~\cite{Kharzeev:2003wz} (KKT04) for 
forward rapidities at RHIC and mid-rapidity at the LHC.

However,  collinear factorized approaches with DGLAP evolved nPDF 
appear to provide an alternate explanation of forward 
single inclusive yields at RHIC ~\cite{EPS08,Eskola:2010jh}.
At LHC energies we can differentiate between these 
explanations because 
the collinear factorized approach predicts only small
nuclear modification for mid-rapidity pions (see Fig.~\ref{fig:fig3} below), 
while at RHIC energies it predicts large modifications, as CGC does,  
for forward produced pions.
The much higher
energy range at the  LHC also opens the kinematic window
on small $x$ physics that can be explored in $p+A$ 
collisions at mid-rapidity.
Some CGC models~\cite{Albacete:2010bs,JalilianMarian:2011dt} predict a 
suppression with  
$R_{p{\rm Pb}}(\eta=0,p_T \approx 10 {\rm GeV}/{\it c}) \approx 0.5$ 
with strong dependence on the initial evolution conditions. 
Such small values of $R_{p{\rm Pb}}$ would 
imply that nearly all nuclear suppression observed in NMF 
$R_{\rm PbPb}$ in Pb+Pb collisions, 
previously attributed to jet quenching in the final state,  
could instead be due to
nonlinear initial state parton flux suppression.

Due to a factor of two increase in the
final parton densities at the LHC, jet quenching is expected to produce higher 
suppression  than at RHIC energies.
Actually, the observed Pb+Pb suppression of pions
at the LHC energy was  surprisingly weaker than expected from RHIC constrained
analysis extrapolated to the LHC~\cite{Horowitz:2011gd}.
Thus, from the perturbative final state interaction point of view
there appears to be no room for initial state suppression. 
Therefore, a measurement of $R_{p{\rm Pb}}$ at mid-rapidity
significantly less than unity would contradict not only perturbative 
QCD (pQCD) models of the
initial state nPDF evolution but also theory of the final state
perturbative opacity series of jet energy loss. 
Since strong coupling AdS/CFT holography 
\cite{Horowitz:2008ig} predicts even stronger final state suppression 
effects, an observation in $p+{\rm Pb}$ of significant deviations from unity 
would then call into question the validity of holographic interpretations
of RHIC and LHC $A+A$ results, including the applicability of minimal
viscous hydrodynamics to apparent perfect fluidity.

At sufficiently high energies and virtualities, QCD factorization theorems
guarantee that jet observables can be calculated in perturbation theory.
The open question is at what scale does 
factorization break down for nuclear processes. 
CGC theory 
\cite{Kharzeev:2003wz,Kharzeev:2004if,JalilianMarian:2011dt,Albacete:2010bs,McLerran:2010ub,Gelis:2010nm,Armesto:2004ud,MC-KLN,Kormilitzin:2010kr,ALbacete:2010ad,Levin:2010zy,Tribedy:2011aa,Dumitru:2011ax}
has a saturation natural scale $Q_s(x,A)$ 
that in principle provides the answer when $Q_s >> \Lambda_{QCD}$. 
However, nuclear jet observables up to LHC energies are sensitive  
to details of  large $x>0.01$ as well as small $A=1$ ``corona'' 
nucleon distributions for which $Q_s \stackrel{<}{\sim} 1 $ GeV.


Monte Carlo models as 
{\small HIJING1.0}~\cite{Wang:1991hta},
{\small HIJING2.0}~\cite{Deng:2010mv} and   
{\small HIJING/B\=B2.0}~\cite{ToporPbPb11} have been developed to study 
hadron productions in $p+p$, $p+A$ and $A+A$ collisions.  
They are essentially two-component models, which describe
the production of hard parton jets and the soft interaction between
nucleon remnants. 
The hard jets production is calculated 
employing collinear factorized multiple minijet within pQCD.
A cut-off scale $p_0$ in the transverse momentum 
of the final jet production has to be introduced below which 
($p_T < p_0$) the
interaction is considered nonperturbative and is characterized by
a finite soft parton cross section $\sigma_{\rm soft}$. 
Jet cross section, depend on the 
parton distribution functions (PDFs) that are parametrized from a 
global fit to data ~\cite{Deng:2010mv}.

Nucleons remnants interact via soft gluon exchanges described by the
string models \cite{Andersson:1986gw,Bengtsson:1987kr} and  
constrained from lower energy $e+e, e+p, p+p$ data.  
The produced hard jet pairs and the two excited remnants
are treated as independent strings, which fragments to resonances that
decay to final hadrons.
Longitudinal beam jet string fragmentations strongly depend on the 
values used for string tensions that control
quark-anti-quark ($q\bar{q}$) and 
diquark-anti-diquark (${\rm qq}\overline{\rm qq}$) pair creation rates
and strangeness suppression factors ($\gamma_s$). 
In the {\small HIJING1.0} 
and {\small HIJING2.0} models a constant (vacuum value) for the effective
value of string tension is used, $\kappa_0 = 1.0$ GeV/fm.
At high initial energy density the novel nuclear physics is due to
the possibility of multiple longitudinal flux tube overlapping 
leading to strong longitudinal color field (SCF) effects.
Strong Color Field (SCF) effects are modeled in {\small HIJING/B\=B2.0} 
by varying the effective string tensions value. 
SCF also modify the fragmentation processes 
resulting in an increase of (strange)baryons which play an important
role in the description of the baryon/meson anomaly.
In order to describe $p+p$ and central Pb + Pb collisions data at
the LHC we have shown that an energy
and mass dependence of the mean value of the string tension  
should be taken into account \cite{ToporPbPb11}.
Moreover, to better describe the baryon/meson anomaly seen in data
a specific implementation of J\=J loops, has to be 
introduced. For a detailed discussion see Ref.~\cite{ToporPbPb11}.   
Similar result can be obtained by including extra diquark-antidiquark
pair production channels from strong coherent fields formed in heavy
ion collisions \cite{peter_2_11}.

All {\small HIJING} type models implement nuclear effects such as nuclear 
modification of the partons distribution functions, i.e., {\em shadowing}
and {\em jet quenching } via 
a medium induced parton splitting process (collisional energy 
loss is neglected)~\cite{Wang:1991hta}. 
In the {\small HIJING1.0} and {\small HIJING/B\=B2.0} models 
Duke-Owen (DO) parametrization of PDFs \cite{DO84} is used to calculate the jet
production cross section with $p_T > p_0$.
In both models using a constant cut-off 
$p_0 = 2$ GeV/{\it c} and a soft parton cross section 
$\sigma_{soft} = 54$ mb fit the experimental $p+p$ data.
However, for $A+A$ collisions in {\small HIJING/B\=B2.0} model  
we introduced  
an energy and mass dependence of the cut-off parameter,
$p_0(s,A)$~\cite{ToporPbPb11} at RHIC and at the LHC energies, 
in order not to violate the geometrical
limit for the total number of minijets per unit transverse area.
In {\small HIJING2.0}~\cite{Deng:2010mv} model 
that is also a modified version of {\small HIJING1.0}~\cite{Wang:1991hta} 
the Gluck-Reya-Vogt (GRV) 
parametrization of PDFs \cite{Gluck:1994uf} is implemented.
The gluon distributions in this different  
parametrization are much higher
than the DO parametrization at small $x$.  
In addition, an energy-dependent cut-off $p_0(s)$ and $\sigma_{soft}(s)$ 
are also assumed in order to better describe the Pb + Pb collisions data at the
LHC.

One of the main uncertainty in calculating  
charged particle multiplicity density
in Pb + Pb collisions is the nuclear modification of parton
distribution functions, especially gluon distributions at small $x$.
In {\small HIJING} type models one assume that 
the parton distributions per nucleon in a nucleus (with atomic number A and 
charge number Z), $f_{a/A}(x,Q^2)$, are 
factorizable into parton distributions in a nucleon ( $f_{a/N}$)
and the parton(a) shadowing factor ($S_{a/A}$),
\begin{equation} 
f_{a/A}(x,Q^2) = S_{a/A}(x,Q^2)f_{a/N}(x,Q^2)
\end{equation}
The impact parameter dependence is implemented through 
the parameter $s_a$,
\begin{equation} 
s_a(b)=s_a \frac{5}{3} \left(1-\frac{b^2}{R_A^2}\right)
\end{equation}
where $R_A = 1.12 A^{1/3}$ is the nuclear radius.

In {\small HIJING/B\=B2.0} the shadowing factor for gluon and quark 
are assumed to be equal ($S_{g/A} (x,Q^2) = S_{q/A} (x,Q^2)$) 
and are similar with those used in {\small HIJING1.0}~\cite{Wang:1991hta}.
They were selected 
in order to fit the centrality dependence of the central 
charged particle multiplicity density at the LHC. 
In contrast, in {\small HIJING2.0} a much stronger impact
parameter dependence of the gluon ($s_g=0.22-0.23$) 
and quark ($s_q=0.1$) shadowing factor is used in order 
to fit the LHC data. 
Due to this stronger gluon shadowing the jet quenching effect has to be
neglected~\cite{Deng:2010mv}.
Note, all {\small HIJING} type models assume a scale-independent form 
of shadowing parametrization (fixed $Q^2$). This approximation could   
breakdown at very large scale due to dominance of gluon emission
dictated by the DGLAP \cite{parisi_77} evolution equation.
At Q = 2.0 and 4.3 GeV/{\it c}, which are typical scales for mini-jet
production at RHIC and LHC respectively, it was shown that the gluon
shadowing varies approximately by $13\%$ in EPS09 parametrizations 
\cite{EPS09}.

\section{Comparative study of model predictions}

\begin{figure} [h!]

\includegraphics[width=\linewidth]{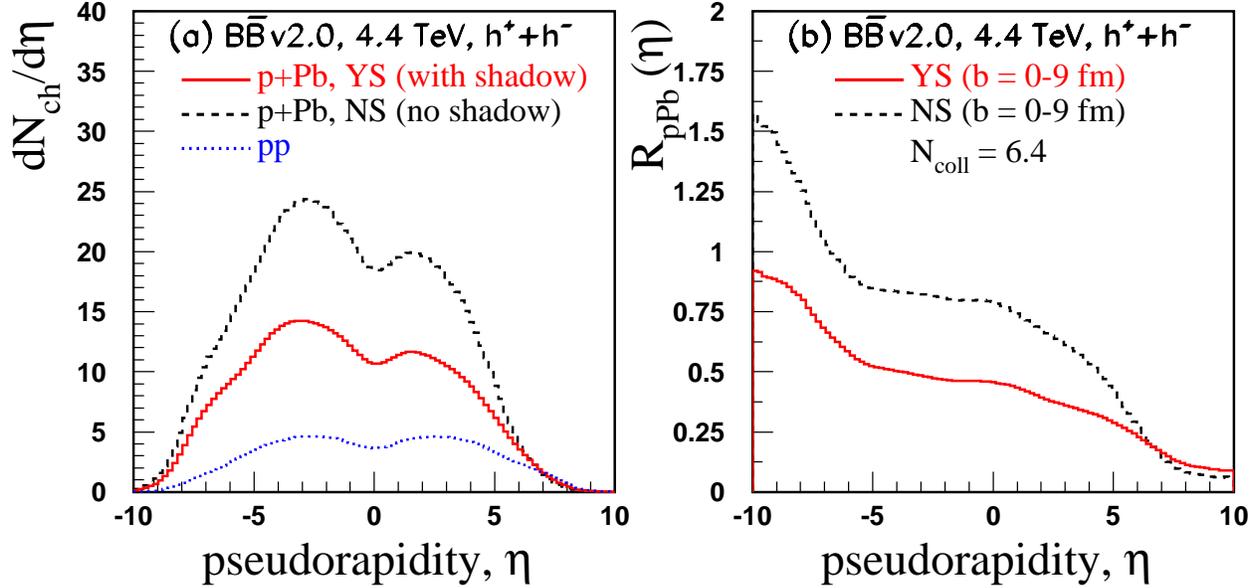}
\vskip 0.5cm
\caption[$p+p$ and $p+Pb$ minimum bias at 4.4 $A$TeV]
{\small (Color online) 
(a) {\small HIJING/B\=B2.0} 
predictions of charged particles pseudorapidity distribution 
$(dN_{\rm ch}/d\eta)$  
for minimum bias (MB) $p$+Pb collisions at 4.4$A$ TeV.   
Solid YS curve includes fixed $Q^2$ shadowing functions  
from {\small HIJING1.0}~{\protect\cite{Wang:1991hta}}, 
while the dashed NS curve has no shadowing.
 (b) Ratio $R_{pPb}(\eta)$ calculated assuming 
$N_{\rm coll}({\rm MB})=6.4$} 
\label{fig:fig1}

\end{figure}

Figure \ref{fig:fig1} shows {\small HIJING/B\=B2.0} 
predictions of the global observables $dN_{\rm ch}/d\eta$ 
and $R_{p{\rm Pb}}(\eta)$ = $(dN_{p{\rm Pb}}^{\rm ch})/d\eta)/
(N_{\rm coll} dN_{pp}^{\rm ch}/d\eta)$ 
characteristics of minimum bias 
$p+{\rm Pb}$ collisions at
4.4$A$ TeV. The predictions for $p+p$ are also shown. 
Minijet cutoff and string tension parameters $p_0=3.1$ GeV/{\it c} 
and $\kappa=2.9$ GeV/fm for  $p+{\rm Pb}$ are determined from
fits to $p+p$ and $A+A$ systematics from RHIC to the LHC 
(see Ref.~{\protect\cite{ToporPbPb11}} for details).
Note, these calculations assume no {\em jet quenching}. 

The absolute normalization of $dN_{\rm ch}/d\eta$ is however sensitive
to the low $p_T < 2$ GeV/{\it c} nonperturbative hadronization
dynamics that is performed via LUND \cite{Andersson:1986gw} 
string JETSET \cite{Bengtsson:1987kr} fragmentation as
constrained from lower energy $e+e, e+p, p+p$ data.  The default
{\small HIJING1.0} parametrization of the fixed $Q_0^2=2$ GeV$^2$ shadow
function leads to substantial reduction (solid histograms) of the
global multiplicity at the LHC. It is important to emphasize that the
no shadowing results (dashed curves) are substantially reduced in 
{\small HIJING/B\=B}2.0 relative to no shadowing prediction with default
 {\small HIJING/1.0} from Ref.~{\protect\cite{Wang:1991hta}},
 because both the default minijet cut-off $p_0=2 $ GeV/{\it c}
and the default vacuum string tension $\kappa_0=1 $ GeV/fm 
(used in {\small HIJING1.0}) are generalized
to vary monotonically with centre of mass (cm) energy per nucleon $\sqrt{s}$ 
and atomic number, $A$.
As discussed in \cite{ToporPbPb11}, systematics of $p+p$ 
and Pb+Pb multiparticle production from RHIC to the LHC 
are used to fix the energy ($\sqrt{s}$) and the $A$ dependence.
Thus the cut-off parameter $p_0(s,A) = 0.416  \; \sqrt{s}^{0.191}
\; A^{0.128}$ GeV/{\it c} and the
mean value of the string tension 
$\kappa(s,A) = \kappa_0 \; (s/s0)^{0.06}\;A^{0.167}$ GeV/fm. 
The above formulae lead to $p_0 = 3.1$ GeV/{\it c} and 
$\kappa = 2.9$ GeV/fm at 4.4$A$ TeV for $p+{\rm Pb}$ collisions.
For $p+p$ collisions at 4.4 TeV we use a constant cut-off parameter
$p_{0pp} = 2 $ GeV/{\it c} and a string tension value 
of $\kappa_{pp} = 2.7$ GeV/fm.

Note, even in the case of no
shadowing shown in Fig.~\ref{fig:fig1}, the increase to $p_0=3.1$
GeV/{\it c} from $p_0 = 2 $ GeV/{\it c} (value used in
$p+p$ at $4.4$ TeV) causes a significant reduction by a factor of
roughly two of the  
minijet cross section and hence final pion multiplicity.  This
reduction of minijet production also is required to fit the low 
charged particle multiplicity growth in $A+A$ collisions 
from RHIC to LHC (a factor of 2.2)~\cite{jharris11}.  

\begin{figure} [h!]

\centering

\includegraphics[width=\linewidth]{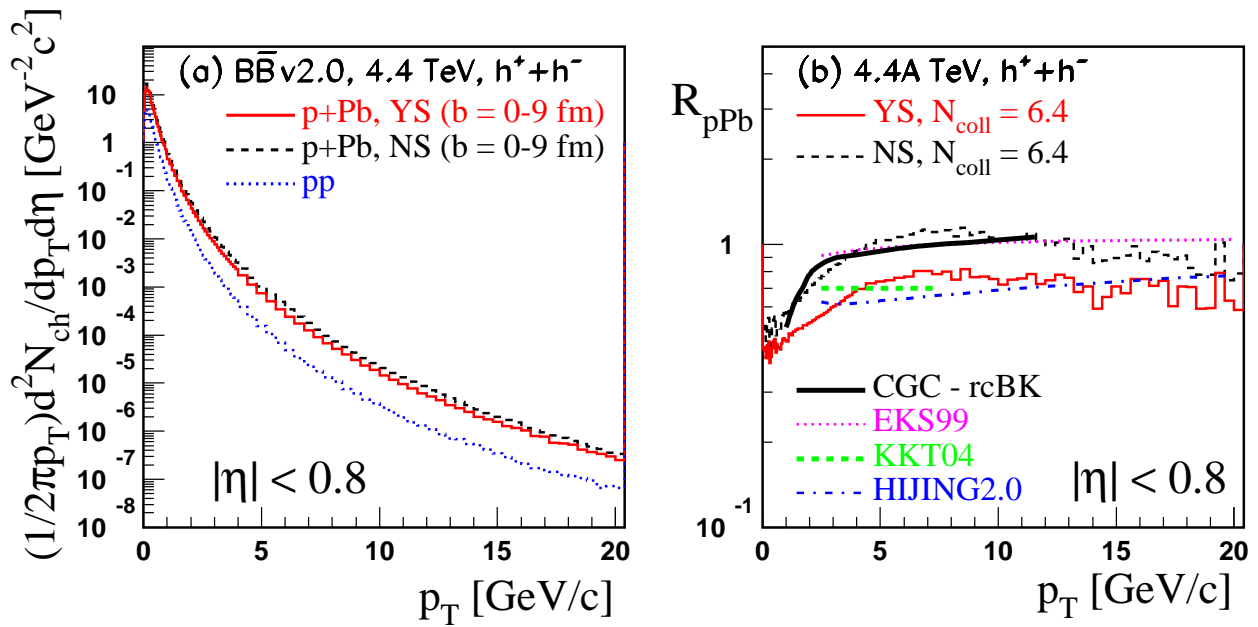}
\vskip 0.5cm\caption[Charge hadron nuclear modification 
factor $p+Pb$ minimum bias at 4.4 $A$TeV]
{\small (Color online) (a) Minimum bias transverse momentum
distributions at mid-pseudorapidity $|\eta|<0.8$
predicted by {\small HIJING/B\=B2.0} with (solid histogram) 
and without (dashed histogram) {\small HIJING1.0} shadowing
functions {\protect\cite{Wang:1991hta}}. The results
for $p+p$ collisions at 4.4 TeV (dotted histogram) are also included.
(b) The mid-pseudorapidity nuclear modification factor of charged hadrons
$R_{p{\rm Pb}} $ from {\small HIJING/B\=B2.0} model.
The solid and thin dashed histograms have the same meaning as in 
part (a). They are compared to pQCD leading order (LO) predictions 
(dash dotted) {\protect\cite{Levai11}}   
using {\small HIJING2.0} shadowing functions~{\protect\cite{Deng:2010mv}},   
and to DGLAP $Q^2$ evolved nPDF, 
EKS99 (dotted)~{\protect\cite{Eskola:1998df}}. 
Predictions of CGC model (thick dashed)~{\protect\cite{Kharzeev:2003wz}} 
(KKT04)  and CGC - rcBK model 
(thick solid) from Ref.~\cite{Tribedy:2011aa} are also included.  
}
\label{fig:fig2}

\end{figure}

We interpret this as  
additional phenomenological evidence for gluon saturation physics not
encoded in leading twist shadow functions. The $p_T>5 $ GeV/{\it c}
minijets tails
are unaffected but the bulk low $p_T<5$ GeV/{\it c} multiplicity distribution 
is sensitive to this extra energy $(\sqrt{s})$ and $A$ dependence of
the  minijet shower suppression effect.
It is difficult to relate $p_0$ to saturation scale $Q_{sat}$ directly, because
in {\small HIJING} hadronization proceeds through longitudinal field
string fragmentation. The energy $(\sqrt{s})$ and $A$ dependence 
of the string tension value 
arises from strong color field (color rope) effects not considered 
in CGC phenomenology that assumes $k_T$ factorized gluon fusion 
hadronization.
{\small HIJING} hadronization of minijets is not via independent 
fragmentation functions
as in PYTHIA \cite{Bengtsson:1987kr}, but via string fragmentation 
with gluon minijets represented as kinks in the strings. The interplay
between longitudinal string fragmentation dynamics and minijets
is a nonperturbative feature of {\small HIJING} type models.  The 
approximate triangular (or trapezoidal) rapidity asymmetry 
seen in the ratio $R_{p{\rm Pb}}(\eta)$ sloping downwards from the
nuclear beam fragmentation region at negative
pseudorapidity $\eta < -5$ toward $1/N_{\rm coll}$ in the proton 
fragmentation region ($\eta > 5$) is a basic Glauber geometric
effect first explained in Refs.~\cite{Brodsky:1977de,Adil:2005qn}
and realized via string fragmentation in {\small HIJING}.

In Fig.~\ref{fig:fig2} are displayed the predicted transverse spectra and
nuclear modification factor for charged hadrons
at mid-pseudorapidity, $|\eta|< 0.8$. 
Including shadowing reduces $R_{p{\rm Pb}}$ from unity
to about 0.7 in the interesting 5 to 10 GeV/{\it c} region close to the 
prediction of Color Glass Condensate
model~\cite{Kharzeev:2003wz} (KKT04). A similar nuclear modification factor
is found  {\protect\cite{Levai11}} using leading order (LO) pQCD 
collinear factorization with
{\small HIJING2.0} parameterization of shadowing functions 
{\protect\cite{Li:2001xa}}, 
GRV parton distribution functions (nPDF) from 
Ref.~{\protect\cite{Gluck:1994uf}}, 
and hadron fragmentation functions from Ref.~{\protect\cite{Kniehl:2000fe}}. 

In stark contrast to the three curves near 0.7$\pm 0.1$ from 
completely different dynamical modeling the standard DGLAP evolved 
global e+A fit
nPDF (dotted curve labelled EKS99 {\protect\cite{Eskola:1998df}}) 
predicts near unity for transverse momenta 
above 5 GeV/{\it c}. 
The no shadowing {\small HIJING/B\=B2.0} values 
(NS, thin dashed histogram) goes to unity above 5 GeV/{\it c}, but the 
nonperturbative string hadronization
pulls the intercept at $p_T=0$ near to $1/2$ as constrained by the 
global triangular enhanced form of $dN_{p{\rm Pb}}/d\eta$ relative 
to $dN_{pp}/d\eta$ shown in Fig.~\ref{fig:fig1}b. 
Note, that the model BGK77 from
Refs.~\cite{Brodsky:1977de,Adil:2005qn}
also predicts $R_{pA}(y,p_T =0) = 1$ at the
nuclear target rapidity
and  $1/N_{\rm coll}(b)$ at the proton projectile rapidity.

However, a recent new version of the CGC-rcBK model~\cite{Tribedy:2011aa}
predicts essentially no shadowing/saturation effects at $\eta=0$ in contrast to
both CGC-KKT04 ~\cite{Kharzeev:2003wz} and  CGC-rcBK model from 
Ref.~\cite{JalilianMarian:2011dt}. 
The absence of shadowing at midrapidity in the
CGC-rcBK~\cite{Tribedy:2011aa}  model
is due to a phenomenological  extra anomalous dimension $\gamma$,
introduced to modify color dipole cross section 
$\sigma_{dipole}(r) \propto (r^2)^\gamma$.
  This  significantly steepens the $pp$ transverse momentum
distribution relative to the quadratic form $\sigma_{dipole}(r) \propto r^2$ 
used in CGC model (MV) \cite{McLerran:1993ni}
as required to reproduce LHC $pp$ data. 
Recently, posible extra $A$ dependence of this extra anomalous
dimension has been proposed~\cite{Dumitru:2011ax}.
It would be very surprising indeed if future $p+Pb$ data would show no
evidence
of shadowing with a $R_{pPb}\approx 1.0$ at  $\eta=0$
mid-pseudorapidity 
which could then be ascribed either  to
(i)  rapid DGLAP $Q^2$ evolution of shadowing in EKS09 \cite{EPS09}
parametrization or to
(ii) accidental cancellation of deep saturation effects due to an anomalous
short distance behavior of the dipole cross section in CGC modeling.

\begin{figure}[h!]
\includegraphics[width=\linewidth]{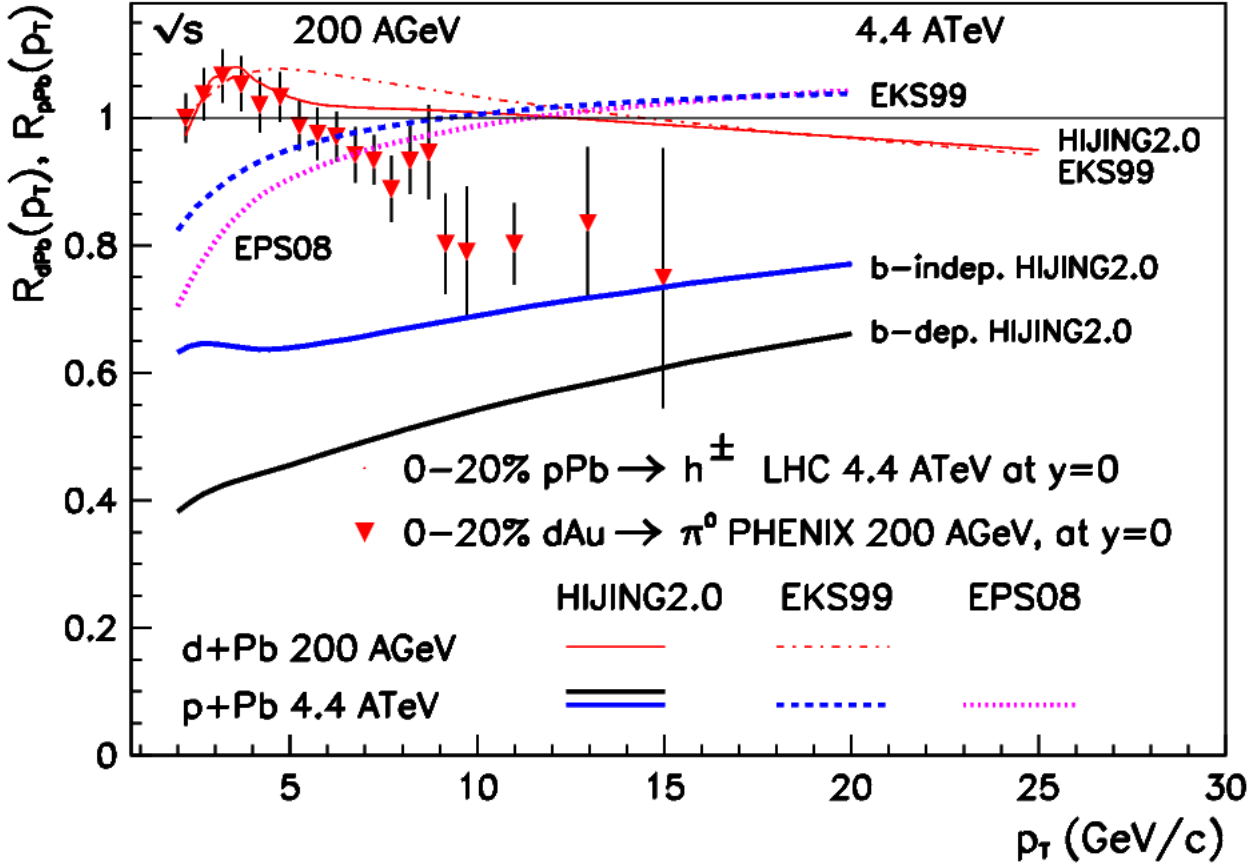}
\vskip 0.5cm 
\caption[Gergely]
{\small (Color online) Predictions  updated 
at 4.4$A$ TeV of Refs.~\cite{Levai11} results for central $0-20\%$ 
($b<3.5$ fm) $p+Pb$ at mid-rapidity. 
The original predictions at 0.2$A$ TeV for $d+{\rm Au}$ are also included.
Compared are the results obtained with fixed $Q^2$ shadowing 
functions {\small HIJING2.0}~{\protect\cite{Deng:2010mv}} 
 with (b-dep) and without (b-indep) impact parameter dependence. 
Predictions with DGLAP $Q^2$ evolved shadowing functions  
from Ref.~{\protect\cite{Eskola:1998df}} (EKS99) and 
Ref.~{\protect \cite{EPS08}} (EPS08) are also shown. 
The data are from PHENIX Collaboration~\cite{Adler:2006wg}.}
\label{fig:fig3}
\end{figure}

What is significant about higher $p_T$ deviations from unity
is that the nuclear modification factor in central $A+A$ collisions
is related to the minimum bias $p+A$ by simple Glauber geometric
considerations, that can be expressed as

\begin{equation}
R_{AA}(y=0,p_T, b=0) \approx \langle (R_{pA}(y=0,p_T, 
{\rm min.bias})) \rangle^2
\end{equation}

where the average is calculated over all impact parameters.
Thus, $R_{p{\rm Pb}} \approx 0.7$ for minimum bias collisions 
implies a NMF $R_{\rm PbPb} \approx 0.5$ in central 
Pb+Pb collisions before any final state interactions take place.
The result for $R_{p{\rm Pb}}(p_T) \approx 0.7$ is similar with those 
reported recently with CGC type models
~\cite{Albacete:2010bs,JalilianMarian:2011dt} (rcBK),
albeit with huge error bars at mid-rapidity because of poorly
known initial saturation conditions
for $p+p$ and near the surface of heavy nuclei. 
However, if the prediction of 
$R_{p{\rm Pb}}\approx 0.5$~\cite{Albacete:2010bs} 
turned out to be confirmed by the upcoming $p+Pb$ measurements 
then, in central Pb+Pb collisions we expect 
a factor of roughly four suppression  ($\approx \,0.25$)
in pions at transverse momenta of roughly 10 GeV/{\it c}. 
This fact would leave no room for final state interactions in matter
100 times denser than ground state nuclei. 
Needless to say this point alone underlines more the importance of 
measuring the $p_T$ dependence of NMF ($R_{p{\rm Pb}} (p_T)$) at the LHC.

In Figure~\ref{fig:fig3} the updated predictions 
at 4.4$A$ TeV~\cite{Levai11} of $R_{p{\rm Pb}}(p_T)$
in central($0-20\%$;$b<3.3$ fm) $p+{\rm Pb}$ collisions at
mid-rapidity are shown. 
The message is similar to that obtained from Fig.~\ref{fig:fig2}.
The standard collinear $Q^2$ evolved nPDF models  
from Ref.~\cite{Eskola:1998df} (EKS99)
and from Ref.~\cite{EPS08} (EPS08)  
predict only a slight deviation ($\approx \,10\%$) from unity, 
as discussed in detail in Refs.~\cite{pALHC11,Wiedemann10}. 
Fixed $Q^2$ shadowing functions used in {\small HIJING1.0} 
or {\small HIJING2.0} models
predict $R_{p{\rm Pb}}(p_T) = 0.6 \pm 0.1$ in the $p_T$ range 
5 to 15 GeV/{\it c}, well below unity. 
Previous results at the RHIC energy ~\cite{Levai11} for 
central ($0-20\%$) $d+{\rm Au}$ collisions at 0.2$A$ TeV
are also presented in comparison with PHENIX data \cite{Adler:2006wg}.
At RHIC energy (0.2$A$ TeV) all models predict approximately 
$R_{d{\rm Pb}}(p_T) = 1 \pm 0.1$. 
At this energy {\small HIJING2.0} the  
shadowing is much weaker for $p_T>5$ GeV/{\it c} domain 
because this correspond to $x > 0.05$, which is over an order of magnitude 
larger than at the LHC energy. 
Taking an impact parameter dependence of shadowing
function in {\small HIJING2.0}~\cite{Deng:2010mv} 
for central $p+{\rm Pb}$ collisions  
results in a further decrease of the NMF $R_{p{\rm Pb}}(p_T)$ by $15-20\%$.

\begin{figure} [h!]

\centering

\includegraphics[width=\linewidth]{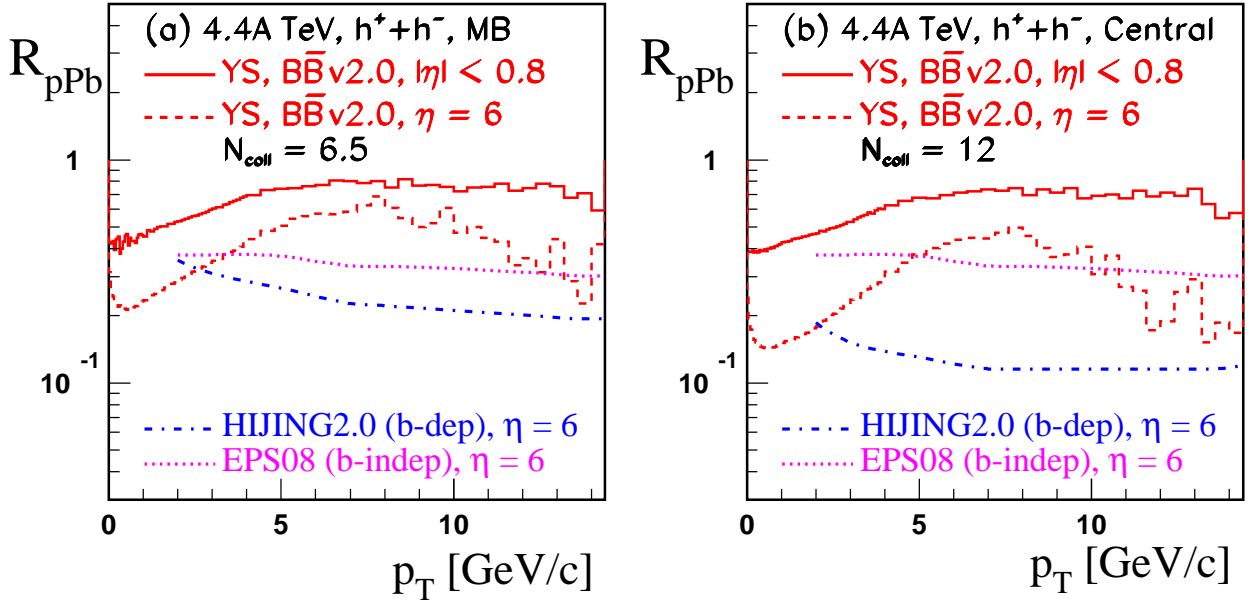}
\vskip 0.5cm\caption[Charge hadron nuclear modification 
factor $p+Pb$ minimum bias at 4.4 $A$TeV]
{\small (Color online) 
(a) The minimum bias (MB) NMF  
of charged particles at forward pseudorapidity 
$\eta = 6$, from {\small HIJING/B\=B2.0} model (dashed histogram).
The results are obtained with shadowing functions from the 
{\small HIJING1.0} model~{\protect\cite{Wang:1991hta}}.  
They are compared to pQCD leading order (LO) results at $\eta = 6$ 
(dash dotted)~{\protect\cite{Levai11}}   
using impact parameter dependent (b-dep) 
{\small HIJING2.0} shadowing functions~{\protect\cite{Deng:2010mv}} and to
predictions obtained with DGLAP $Q^2$ evolved shadowing functions 
(dotted) with no impact parameter dependence (b-indep) 
from Ref.~\cite{EPS08}.
For reference the results at mid-pseudorapidity, $|\eta|<0.8$
(solid histogram) are also included.
(b) The results obtained for NMF 
of charged particles in  central ($N_{\rm coll} = 12 $)
$p$+Pb collisions at 4.4$A$ TeV. 
The histograms and the lines have the same meaning as in part (a).}

\label{fig:fig4}

\end{figure}

Finally, in Fig.~\ref{fig:fig4} we 
show the nuclear modification factor $R_{pPb}$ for inclusive charged
hadrons ($h^+ + h^-$) at $\sqrt{s} = 4.4$ TeV obtained from
different models for minimum bias (MB) ($N_{\rm coll}$ = 6.4) 
in Fig.~\ref{fig:fig4}a and central ($N_{\rm coll}$ = 12)
$p$ + Pb collisions in Fig.~\ref{fig:fig4}b at forward pseudorapidity
($\eta = 6 $).
For reference we include the NMF at mid-pseudorapidity
$|\eta|<0.8$ (solid histograms) predicted with 
{\small HIJING/B\=Bv2.0}.  
Smaller suppression in the region of interest ($5 < p_T < 10$ GeV/c) 
than obtained within {\small HIJING/B\=B v2.0}
model is predicted using EPS08 parametrization \cite{EPS08} with
no impact parameter dependence (b-indep) of shadowing functions (dotted) 
and with pQCD calculation 
using an impact parameter dependence (b-dep) from {\small HIJING2.0} model. 
In this range of transverse momenta, 
the {\small HIJING/B\=B v2.0} model predict slightly higher values
than predicted (0.35 -0.40) by 
CGC model (rcBK)~\cite{JalilianMarian:2011dt}. 

It is obvious that at forward pseudorapidity the suppression 
is higher for central than in minimum bias (MB) $p$ + Pb collisions.
Moreover, for central $p$ + Pb collisions the sensitivity to 
parametrization of shadowing functions is amplified.
The different shape predicted by {\small HIJING/B\=B v2.0} 
at forward pseudorapidity in both minimum bias (MB) and central collisions
could be explained as a specific interplay between J\=J loops and 
SCF effects embedded in the model, which induce a baryon/meson
anomaly. 
Note, the same effect has been predicted in $p+p$ and Pb+Pb collisions
at LHC energies~\cite{ToporPbPb11}.
To draw a definite conclusion, measurements of identified  
particle NMF, $R_{p{\rm Pb}}^{\rm ID}(p_T)$ are needed.
Such measurements will provide a vital
information on cold nuclear matter effects and will 
constrain the main parameters of shadowing functions 
used within different models.

\section{Conclusion}

In conclusion, even with a small sample of $10^6$ events 
the study of $R_{p{\rm Pb}}(p_T)$ or central relative to peripheral NMF
($R_{\rm CP}(p_T)$) could provide a definitive constraint on nuclear 
shadowing implemented within different pQCD inspired models and CGC saturation
models, with high impact on the interpretation or reinterpretation of the 
bulk and hard probes for nucleus-nucleus (Pb+Pb)
collisions at LHC energies.

{\small HIJING} type models predict 
for central ($N_{\rm coll} = 12$) $p+{\rm Pb}$ collisions  
smaller values of nuclear modification factors ($R_{p{\rm Pb}}(p_T)$) 
than in minimum bias  events. 
The possibility to trigger on
the highest multiplicity tails of 
transverse momentum spectra in $p+{\rm Pb}$ collisions, 
will open the way to study collective phenomena  
in proton nucleus interactions with superdense nuclear cores, where 
average number of binary collisions could increase to 
$N_{\rm coll} > 12$. These measurements will provide an  
stringent test of the phenomenological models discussed in this paper.

\section{Acknowledgments}
\vskip 0.2cm 

{\bf Acknowledgments:} Discussion with 
B.~Cole, J.~Harris, G.~Roland, J.~Schukraft, and W.~Zajc 
as well as B.~Mueller, A.~Dumitru, 
Jamal Jalilian-Marian, D.~Kharzeev, L.~McLerran, and Xin-Nian Wang  are 
gratefully acknowledged.
VT and JB are  supported by the Natural Sciences and Engineering 
Research Council of Canada.  
MG is supported by the Division of Nuclear Science, 
U.S. Department of Energy, under Contract No. DE-AC03-76SF00098 and
DE-FG02-93ER-40764 (associated 
with the JET Topical Collaboration Project). 
GGB, MG, and PL  also thanks for the Hungarian grants OTKA PD73596, NK778816, 
NIH TET\_10-1\_2011-0061 and ZA-15/2009. 
GGB was partially supported by the J\'anos Bolyai Research Scholarship 
of the HAS.

\end{document}